\documentclass[a4paper,11pt]{article}
\pdfoutput=1 
\usepackage{jcappub}

\usepackage[T1]{fontenc}
\usepackage{xcolor}
\usepackage{hyperref}
\usepackage[figuresright]{rotating}
\usepackage{graphicx}
\usepackage[utf8]{inputenc}
\usepackage{color}
\usepackage[normalem]{ulem}
\usepackage{amsmath}

\title{\boldmath Correlation Analysis of Decaying Sterile Neutrino Dark Matter in the Context of the SRG Mission}

\author{V.~V.~Barinov}

\affiliation{Institute for Nuclear Research of the Russian Academy of Sciences, Moscow 117312, Russia}
\affiliation{Physics Department, Lomonosov Moscow State University, Leninskie Gory, Moscow 119991, Russia}

\emailAdd{barinov.vvl@gmail.com}

\abstract{We provide a correlation analysis of signatures associated with traces of the dark matter decay and the galaxy spatial distribution according to the 2MRS catalog of galaxies. Signature data analysis plays an important role in the context of current and future observations and cosmological constraints. Attention is paid to the constraints that can be obtained for decaying sterile neutrinos when analyzing observations in the context of the Spectr-Roentegn-Gamma (SRG) mission. We study the correlation spectra of dark matter and galaxies, which can be obtained both for the eROSITA telescope and for the first time for the ART-XC telescope. The analysis is carried out both within the framework of the Limber approximation and within the framework of the extended Limber approximation, which makes it possible to more accurately study the power spectra in the region of small multipoles. We calculate the power spectra in both approaches and examine the contribution of different ranges of multipoles to the resulting constraints on sterile neutrino parameters.}

\begin{document}
\maketitle
\flushbottom

\section{Introduction}\label{sec:intro}
The question of the nature of dark matter is one of the most pressing issues of fundamental physics. As is known, dark matter manifests itself only through gravitational interaction, and to date there is no sufficiently confirmed evidence that reveals the non-gravitational nature of dark matter. Nevertheless, an active search for dark matter particles continues, which can potentially be detected outside the framework of only gravitational interaction.

In this regard, sterile neutrinos with masses of several keV seem to be quite promising for the role of dark matter particles~\cite{Adhikari:2016bei, Boyarsky:2018tvu}. Sterile neutrinos of given masses can exhibit properties of both cold (CDM) and warm dark matter (WDM)~\cite{Abazajian:2001vt, Abazajian:2001nj}, which potentially circumvents some of the problems inherent in models with only cold dark matter (CDM)~\cite{Dodelson:1993je, Shi:1998km, Shaposhnikov:2006xi, Lovell:2013ola, Bezrukov:2014nza, Adhikari:2016bei, Bezrukov:2018wvd}.

In this paper, we focus on a specific dark matter candidate, namely sterile neutrinos that are unstable due to mixing with active neutrinos and therefore decay into an active neutrino (electron, muon or tau neutrino) and a photon,
\begin{equation*}
  \label{decays}
\nu_{s} \rightarrow \nu_{e, \mu, \tau} + \gamma.
\end{equation*}
The sterile neutrino decay width in this process is given by the following expression ~\cite{PhysRevD.25.766, Barger:1995ty}
\begin{equation}\label{eq:neutrino_width}
\nonumber\Gamma_{\nu_s} = \frac{9}{1024} \frac{\alpha}{\pi^4} G_F^2 m_{\nu_s}^5\sin^22\theta = 1.36 \times 10^{-22}\left(\frac{m_{\nu_s}}{1 \text{keV}}\right)^5 \sin^22\theta\hspace{0.25cm} \text{s}^{-1},
\end{equation}
where $m_{\nu_s}$ is the sterile neutrino mass, $\theta$ is the mixing angle between the active and sterile neutrinos, where we do not distinguish between the mass and flavor eigenstates at small mixing angles. In this a two-body decay, the energy of the outgoing photon is $E_{\gamma}=m_{\nu_s}/2$, and the sterile neutrinos that form the galactic dark matter produce a monochromatic photon spectrum with a width of the order of the speed of dark matter particles in the galaxy, i.e. $v\sim10^{-4}\!-\!10^{-3}$.

There have been many works devoted to the detection of traces of sterile neutrino decay from astrophysical observations~\cite{Abazajian:2001vt, Boyarsky:2005us, Boyarsky:2006fg, Boyarsky:2006kc, Boyarsky:2014ska, Ruchayskiy:2015onc, Yunis:2018eux, Perez:2016tcq, Ng:2019gch, Dekker:2021bos, Roach:2022lgo} partly stimulated by the announced hints of positive results~\cite{Bulbul:2014sua, Boyarsky:2014jta}. The detection of a signal from decay traces is usually carried out observing the X-ray background of the blank sky field or when observing galaxies and galaxy clusters in a given direction. Then, the expected signal from dark matter in a given direction is calculated for a given dark matter density distribution profile, and the signal/background ratio is examined to obtain constraints on the parameters of sterile neutrinos. The latest results severely constrain the parameter space of decaying sterile neutrinos~\cite{Perez:2016tcq, Ng:2019gch, Roach:2022lgo}.

Constraints on the parameters of decaying sterile neutrinos can be obtained not only within the framework of direct astrophysical observations of galaxies, galaxy clusters, and parts of the blank sky field. It is also possible to constraint the parameter space of decaying (annihilating) dark matter within the framework of the correlation analysis of various signatures~\cite{Cooray:2002dia, Fornengo:2013rga},~\cite{ Zandanel:2015xca, Caputo:2019djj},~\cite{2009MNRAS.400.2122A, Ando:2013xwa, Ando:2014aoa, Ando:2017wff, Cuoco:2015rfa}. 

Indeed, since dark matter particles are concentrated inside galaxies and galaxy clusters, each photon from the decay of dark matter should point to a specific object where the decay occurred. If the photons were not deflected, then they indicate part of this object in the sky. Including the spatial distribution of this object due to the redshift. Even if the object cannot be recognized by the observer until now (unresolved sources), then the connection between the photon and its source exists and can be traced statistically, by joint analysis of the distribution of all registered photons in the arrival direction, energy and distribution map of cosmic structures. Note that galaxies and galaxy clusters are sources of X-ray radiation of astrophysical origin, uncorrelated with decaying dark matter. However, a correlation between observed photons and cosmic structure must exist anyway, even without any contribution from the decaying of dark matter.

This approach is based on the study of the auto and cross-correlation angular power spectrum of dark matter and galaxies. As a part of the correlation analysis, a nonlinear power spectrum is calculated for each pair of signatures (dark matter - dark matter, galaxies - galaxies, dark matter - galaxies) and then a cross-correlation function is constructed for all pairs of signatures under study. This angular correlation power spectrum is obtained by integrating over the redshift ($z$) and the conformal momentum ($k$) of the nonlinear power spectrum and the window function describing the characteristic shape of the tracer distribution over the sky. Then the resulting model spectrum is compared with the measured cross-correlation spectrum, taking into account the uncertainties from the X-ray background and contribution of the uncertainties from given signatures. Thus, we are looking for correlations between the power spectra of such structures as galaxies, clusters of galaxies throughout the sky at various redshifts, and the power spectrum of dark matter due to the decay of sterile neutrinos. Within the framework of this approach, it becomes possible to study the cross-correlation angular power spectrum, the anisotropy of the signal from dark matter on various cosmological scales.

In this paper we follow the analysis of ~\cite{Zandanel:2015xca, Caputo:2019djj} within the approach described above. We calculate angular correlation spectra of dark matter due to the decay of sterile neutrinos into active neutrinos and photons, correlation spectra for the catalog of galaxies, and cross-correlation spectra of dark matter and the catalog of galaxies. We use the 2MRS~\cite{Huchra_2012} catalog as the base catalog, which covers most of the sky and which was also used in previous work~\cite{Ando:2017wff}. Based on these data, we constrain the space of parameters of sterile neutrinos. We perform calculations for both the eROSITA~\cite{2010SPIE.7732E..0UP,2012arXiv1209.3114M,erosita2020} and ART-XC~\cite{2011SPIE.8147E..06P,Pavlinskypart1, Pavlinskypart2,2019ExA....48..233P,Pavlinsky:2021jsk, Pavlinsky:2021ynp} telescopes using X-ray background estimates from these telescopes~\cite{Predehl:2020waz, Pavlinsky:2021jsk, Pavlinsky:2021ynp, Barinov:2020hiq}. We do not limit ourselves to the standard approach within the Limber approximation~\cite{1953ApJ...117..134L} but also carry out correlation analysis using the extended Limber approximation~\cite{LoVerde:2008re} and then compare the results obtained within the both approaches. We show that the constraints obtained by us are consistent with the results of the previous works~\cite{Zandanel:2015xca, Caputo:2019djj}. Additionally, it is shown that the constraints are accumulated mainly at average and high multipoles. We perform a combined constraint analysis within the correlation framework and show that the constraints that can be derived from this approach remain weaker than those that can be derived from direct astrophysical observations.

The work is organized as follows. In section~\ref{sec:formalism}, we outline the general formalism for calculating correlation spectra. In sections~\ref{subsec:limber} and~\ref{subsec:ExtendLimber}, we depict the Limber and extended Limber approximations. In sections~\ref{sec:auto-corr-spectrum},\ref{sec:auto-corr_gal},\ref{sec:cross_corr_dmgal} an approach to the calculation of correlation spectra is explained. In section~\ref{sec:exp_setup}, we introduce the main features of the eROSITA and ART-XC telescopes. In section~\ref{sec:space_cons} we describe the analysis procedure and the results obtained. In section~\ref{sec:conclusion} we make a conclusion.

\section{General Formalism}\label{sec:formalism}
The correlation function of intensity fluctuations of the signatures $i$, $j$ is defined as~\cite{1980lssu.book.....P, Cooray:2002dia, Weinberg:2008zzc, Gorbunov:2011zzc, Fornengo:2013rga}
\begin{equation}
    \langle \delta I_i(\vec{n_1}) \delta I_j(\vec{n_2}) \rangle = \sum_{l}\frac{2l+1}{4\pi}C_{l}^{ij}P_{l}(cos(\theta)),
\end{equation}
where by signatures we mean the emission of X-ray photons caused by the decays of dark matter particles and the distribution of the number of galaxies depending on the redshift, $\vec{n}_{1}$, $\vec{n}_{2}$ are the unit direction vectors on the celestial sphere, and the angle $\theta = \angle (\vec{n}_1, \vec{n}_2)$ corresponds to the angular size of the given part of the sky, $C_{l}^{ij}$ is the angular power correlation spectrum between signature fluctuations $i$ and $j$, $P_{l}(cos(\theta))$ are the Legendre polynomials, $\delta I_{i}( \vec{n}) =  I_{i}( \vec{n}) - \langle I_{i} \rangle$ are intensity fluctuations of different signatures, where $\langle I_{i} \rangle$ is the average intensity over the sky.

The angular power correlation spectrum $C_{l}^{ij}$ is the two-point correlation function for the given signatures. It determines the magnitude and properties of signature anisotropy and is given by the following expression~\cite{Fornengo:2013rga}
\begin{equation}
   \label{eq:aps_general}
    C_l^{ij} = \frac{2}{\pi}\ 
                \int_{0}^{\infty}\text{d}\chi \int_{0}^{\infty}\text{d}\chi' \int_{0}^{\infty} k^2\text{d}k  \
                \overline{W}_{i}(\chi)\ \overline{W}_{j}(\chi')\
                P_{ij}\left(k, \chi, \chi' \right)\ 
                j_{l}(k\chi)j_{l}(k\chi'),
\end{equation}
where $\chi$ is the comoving distance, $j_{l}(k\chi) $ is the spherical Bessel function of order $l$, $k$ is the comoving wavenumber, $P_{ij}\left(k, \chi, \chi' \right)$ is the  nonlinear matter power spectrum
\begin{equation}
    P_{ij}\left(k, \chi, \chi' \right) = \sqrt{P_{ij}\left(k, \chi \right)P_{ij}\left(k, \chi' \right)},
\end{equation}
which can be calculated in the Halo Model approach~\cite{Cooray:2002dia}, and $\overline{W}_{i}(z)$ is the cumulative window function in the given energy range
\begin{equation}\label{eq:cwf}
\overline{W}_{i}(z) = \int_{E_{min}}^{E_{max}}\text{d}E W_{i}(E, z),
\end{equation}
where the window function $W_{i}(E, z)$ is the redshift distribution of the observed value associated with the signature under study, depending on the given energy~$E$. If the window function $W_{i}$ does not depend on energy, then the quantity $\overline{W}_{i}$ coincides with $W_{i}$.
\subsection{Power Spectrum}\label{subsec:PowerSpectrum}
The large scale distribution of dark matter turns out to be a complicated structure consisting of many individual substructures. The Halo Model~\cite{Cooray:2002dia} is a phenomenological model that describes the density distribution of dark matter contained in isolated halos. Within the framework of this model, two-point correlation functions of dark matter density fluctuations are considered in two cases: when several dark matter centers belong to the same halo and when dark matter centers belong to different halos. In the latter, the intersection of these two types of halos is not considered. Thus, consideration of the correlation functions of fluctuations in the density of dark matter in different halos makes it possible to obtain an expression for the nonlinear matter power spectrum~\cite{Cooray:2002dia, mo_van_den_bosch_white_2010, Fornengo:2013rga,Massara:2014kba}
\begin{equation}
\label{eq:hm_3dps}
    P_{ij}(k,z) = P_{ij}^{1h}(k, z) + P_{ij}^{2h}(k,z),
\end{equation}
where the first term takes into account the correlations between matter particles (centers) belonging to the same halo, and the second term describes the correlation between matter particles (centers) belonging to different halos,
\begin{equation}
    \label{eq:1hterm}
    P_{ij}^{1h}(k, z) = \int \text{d}M \frac{\text{d}n(M, z)}{\text{d}M}\left(\frac{M}{\overline{\rho}(z)}\right)^2u_{i}^{*}(k; M, z)u_{j}(k; M, z),
\end{equation}
\begin{align}
    \label{eq:2hterm}
    P_{ij}^{2h}(k, z)  = & \left[\int \text{d}M_1 \frac{\text{d}n(M_1, z)}{\text{d}M_1}\left(\frac{M_1}{\overline{\rho}(z)}\right)b_{i}(M_1, z)u_{i}^{*}(k; M_1, z)\right] \nonumber \times \\
                        & \left[\int \text{d}M_2 \frac{\text{d}n(M_2, z)}{\text{d}M_2}\left(\frac{M_2}{\overline{\rho}(z)}\right)b_{j}(M_2, z)u_{j}(k; M_2, z)\right]P_{\text{lin}}(k, z),
\end{align}
where $\text{d}n/\text{d}M(M,z)$ is the halo mass function, which has the meaning of the number of isolated gravitationally bound structures per unit mass per unit comoving volume, $\overline{\rho}(z) = \Omega_{\text{CDM}}(z) \rho_{crit}(z)$ is the comoving density of the dark matter background, $b_{i}(M; z)$ is the linear bias~\cite{2010ApJ...724..878T}, $P_{\text{lin} }(k, z)$ is the linear matter power spectrum, $u_{i}(k; z, M)$ is the Fourier transform of the dark matter density distribution function \cite{Cooray:2002dia}
\begin{equation}
\label{eq:normfourie}
    u(\boldsymbol{k}; M, z) = \frac{\int \text{d}^3 \boldsymbol{r} \rho(\boldsymbol{r}; M, z)e^{-i\boldsymbol{kr}}}{\int \text{d}^3 \boldsymbol{r} \rho(\boldsymbol{r}; M, z)},
\end{equation}
where we put $\boldsymbol{k} = (0, 0, k)$.

In the case of a spherically symmetric density distribution, this expression has the form
\begin{equation}
\label{eq:fourie}
        u(k; M, z) = \frac{1}{M}\int_{0}^{R_{vir}}4\pi r^2\text{d}r \frac{\sin(kr)}{kr}\rho(r; M, z),
\end{equation}
where $R_{vir}$ is the virial radius containing a halo of mass $M$.

We use the NFW~\cite{Navarro:1996gj} profile as the dark matter density distribution function $\rho(r; M, z)$
\begin{equation}
    \rho(r; M, z) = \frac{\delta_c(M, z) \rho_{crit}(z)}{(r/r_s)(1 + r/r_s)^2},
\end{equation}
where the critical density $\rho_{crit}(z)$ is expressed in terms of the Hubble parameter $H(z)$
\begin{equation}
    \rho_{crit}(z) = \frac{3H^2(z)}{8 \pi G},
\end{equation}
and characteristic density is written as~\cite{Navarro:1996gj}
\begin{equation}
    \delta_c = \frac{\Delta(z)}{3} \
    \frac{c(M, z)^3}{\left[\ln{\left[1+c(M,z)\right]} - c(M, z)/(1+c(M, z))\right]}.
\end{equation}
The quantity $\Delta(z)$ characterizes the ratio of the average density of the dark matter halo $\rho_{vir}$ within the virial radius to the $\overline{\rho}$ at given $z$. The relationship between the virial radius $R_{vir}$, the halo mass $M$, and the average density $\rho_{vir}$ is expressed as
\begin{equation}
M = \frac{4 \pi}{3}\Delta(z)\overline{\rho}(z)R_{vir}^3,
\end{equation}
where $\Delta(0) = 18\pi^2$ at $z=0$~\cite{Bryan:1997dn}. Note that there are several different ways to determine this value. We set $\Delta(z) = 200$ for all $z$ values. $c(M, z)$ is defined as the ratio of the halo virial radius to the scaling radius $c(M, z) = R_{vir}/r_{s}$. Thus, the dark matter density distribution given by the NFW profile can be described by the halo mass (or virial radius) and the concentration parameter $c(M, z)$. There are many different parametrizations of $c(M, z)$, which can be obtained in particular using numerical simulations. In this paper, we use the model~\cite{Diemer:2018vmz}.

\subsection{Limber Approximation}\label{subsec:limber}

In the large multipole approximation~\cite{1953ApJ...117..134L} when $l \gg 1$, $l \approx kr = k\chi$, the spherical Bessel functions in the integrals~(\ref{eq:aps_general}) rapidly oscillate and can be replaced here with
\begin{equation}
    j_{l}(kr) \approx \sqrt{\frac{\pi}{2l}}\delta(l-kr), l \gg 1,
\end{equation}
Thus, the angular correlation power spectrum~(\ref{eq:aps_general}) in the large multipole approximation takes the form
\begin{equation}
   \label{eq:aps}
    C_l^{ij} = \int_{0}^{\infty} \ 
    \frac{\text{d}\chi}{\chi^2(z)} \
    \overline{W}_{i}(\chi)\overline{W}_{j}(\chi) \
    P_{ij}\left(k\approx\frac{l}{\chi}, z\right).
\end{equation}

\subsection{Extended Limber Approximation}\label{subsec:ExtendLimber}

The Expression~(\ref{eq:aps}) is used in most cases when it is necessary to estimate the angular power correlation spectrum at large multipoles. As a rule, for $l \sim 100$, the expression agrees very well with what formula~(\ref{eq:aps}) gives. However, the question arises of how it is possible to estimate the magnitude of the angular power correlation spectrum for relatively small multipoles. In this case, the calculation of the power spectrum, according to formula~(\ref{eq:aps_general}), seems to be very expensive from a computational point of view. In this case, we can use the so-called extended Limber approximation~\cite{LoVerde:2008re}. 

According to this approach, the expression~(\ref{eq:aps_general}) can be represented as a series in powers of $(l + 1/2)$
\begin{align}
   \label{eq:aps_general_bessel_series}
    \nonumber C_l^{ij} =& \int_{0}^{\infty} \ 
    \frac{\text{d}\chi}{\chi} \
    \overline{w}_{i}(\chi)\overline{w}_{j}(\chi) \
    P_{ij}\left(k=\frac{\nu}{\chi}, z\right) \times\\ 
    & \left(1-\frac{1}{\nu^2}\left(\frac{\chi^2}{2}\left(\frac{\overline{w}_{i}''}{\overline{w}_{i}}+\frac{\overline{w}_{j}''}{\overline{w}_{j}}\right)+\frac{\chi^3}{6}\left(\frac{\overline{w}_{i}'''}{\overline{w}_{i}}+\frac{\overline{w}_{j}'''}{\overline{w}_{j}}\right)\right) + \mathcal{O}\left(\frac{1}{\nu^4}\right)\right),
\end{align}
where $\nu = l+1/2$, $\overline{w}_{i}(\chi) \equiv \overline{W}_{i}(\chi)/\sqrt{\chi}$, and here we restrict ourselves to the first two terms of the expansion.

Thus, the angular power correlation spectrum can be written as an expansion in powers of $(l+1/2)$ to terms with the required degree of accuracy. Just as in work~\cite{LoVerde:2008re}, we note that even replacing $l \rightarrow l+1/2$ in expression~(\ref{eq:aps}) reduces the error from $\mathcal{O}(l^{-1})$ to $\mathcal{O}(l^{-2})$. This is a very useful result, which makes it possible to improve the estimates obtained within the framework of this approximation quite well. 

\section{Autocorrelation dark matter angular power spectrum}\label{sec:auto-corr-spectrum}

The autocorrelation angular power spectrum of the decaying dark matter has the form, which is determined by equation~(\ref{eq:aps_general}), upon the substitution $P_{ij}(k, z) = P_{dm, dm}(k, z)$ and  $\overline{W}_{i} = \overline{W}_{dm}$ from equation~(\ref{eq:cwf}). In the Limber approximation~\cite{1953ApJ...117..134L}, the expression for this spectrum has the simplest form
\begin{equation}
   \label{eq:aps_dmdm}
    C_l^{dm, dm} = \int_{0}^{\infty} \
    \frac{\text{d}\chi}{\chi^2}
    \overline{W}^2_{dm}(\chi)\
    P_{dm, dm}\left(k=\frac{l}{\chi}, z\right),
\end{equation}
where the nonlinear power spectrum of matter $P_{dm, dm}(k, z)$ is calculated within the framework of the halo model~\cite{Cooray:2002dia, mo_van_den_bosch_white_2010, Fornengo:2013rga,Massara:2014kba}
\begin{equation}
    P_{dm, dm}(k,z) = P_{dm, dm}^{1h}(k, z) + P_{dm, dm}^{2h}(k,z),
\end{equation}
where the terms $P_{dm, dm}^{1h}(k, z)$, $P_{dm, dm}^{2h}(k, z)$ are explicitly written as
\begin{equation}
    P_{\text{dm, dm}}^{1h}(k, z) = \
    \frac{1}{\overline{\rho}^2(0)} \\
    \int \text{d}M \frac{\text{d}n(M, z)}{\text{d}M} \\
    \left[\int_{0}^{R_{vir}}4\pi r^2\text{d}r \frac{\sin(kr)}{kr}\rho(r; M, z)\right ]^2,
\end{equation}
\begin{align}
 \nonumber   P_{\text{dm, dm}}^{2h}(k, z) = &
    \frac{1}{\overline{\rho}^2(0)}
    \left[
    \int \text{d}M \frac{\text{d}n(M, z)}{\text{d}M}b_{\text{lin}}(M, z)\left[\int_{0}^{R_{vir}}4\pi r^2\text{d}r \frac{\sin(kr)} {kr}\rho(r; M, z)\right]
    \right]^2 \times \\
    & P_{\text{lin}}(k, z),
\end{align}
the Window function $W_{\text{dm}}(E, z)$ has the following form~\cite{Zandanel:2015xca}
\begin{equation}
     W_{\text{dm}}(E, z) = \frac{\Omega_{\text{CDM}} \rho_{crit}}{(1+z)}\frac{\Gamma_{\nu_{s}}}{4 \pi m_ {\nu_{s}}}\frac{1}{\sqrt{2 \pi \sigma_{E}^2}} \exp{\left[-\frac{\left(E - \frac{m_{\nu_{s}}}{2(1+z)}\right)^2}{2\sigma_E^2}\right]},
\end{equation}
where $E$ is the energy of the X-ray photon received by the detector, $\sigma_{E} = \text{FWHM(E)} / \sqrt{8\ln2}$ is the energy dispersion of the telescope.

\section{Autocorrelation angular power spectrum of galaxies}\label{sec:auto-corr_gal}
The autocorrelation angular power spectrum of galaxies has the form, which is determined by equation~(\ref{eq:aps_general}) upon the substitution $\overline{W}_{i} = {W}_{g}$, $P_{ij}(k, z) = P_{g, g}(k, z)$. In the Limber approximation~\cite{1953ApJ...117..134L}, the expression for this spectrum has the simplest form
\begin{equation}
   \label{eq:aps_galgal}
    C_l^{g, g} = \int_{0}^{\infty} \
    \frac{\text{d}\chi}{\chi^2(z)} \
    W_{g}^2(\chi)\
    P_{g, g}\left(k=\frac{l}{\chi}, z\right),
\end{equation}
where the nonlinear power spectrum $P_{g, g}(k, z)$ can be calculated within the Halo Occupation Distribution (HOD) formalism~\cite{Seljak:2000gq,Berlind:2001xk,Berlind:2002rn,Cooray:2002dia,Sheth:2002cs,Zheng:2004id,mo_van_den_bosch_white_2010}.

Assuming that the probability distribution function of the presence of $N_{g}$ galaxies in the parent halo with mass $M$ corresponds to the Poisson distribution $P_{\langle N_{g} \rangle_{M}}(N_{g})$~\cite{Zheng:2004id}, the expression for the power spectrum $P_{g, g}(k, z)$ is written as
\begin{equation}
\label{eq:3dps_gg}
    P_{g, g}(k,z) = P_{g, g}^{1h}(k, z) + P_{g, g}^{2h}(k,z),
\end{equation}
\begin{equation}
\label{eq:3dps_gg_1h}
    P_{\text{g, g}}^{1h}(k, z) = \int \text{d}M \frac{\text{d}n(M, z)}{\text{d}M }\frac{\langle N_{g}(N_{g} - 1)\rangle_{M}}{\langle n_g(z) \rangle^2}|u_{g}(k; M, z)|^ p,
\end{equation}
\begin{equation}
\label{eq:3dps_gg_2h}
     P_{\text{g, g}}^{2h}(k, z) = \left[\int \text{d}M \frac{\text{d}n(M, z)}{\text{ d}M}\frac{\langle N_g\rangle_{M}}{\langle n_g(z) \rangle}b_{\text{lin}}(M, z)u_{g}(k; M, z) \right]^2 P_{\text{lin}}(k, z).
\end{equation}
Due to the fact that the second term makes a significant contribution to the power spectrum mainly on large scales, the expression~(\ref{eq:3dps_gg_2h}) with normalization $u(k; M, z) \rightarrow 1$ for $k \rightarrow 0$ is approximately equal to
\begin{equation}
    P_{\text{g, g}}^{2h}(k, z) \approx b_{g}^{2}(z)P_{\text{lin}}(k, z),
\end{equation}
where $b_{g}$ is the galactic bias~\cite{Cooray:2002dia}
\begin{equation}
     b_{g}(z) = \frac{1}{\langle n_g(z) \rangle}\int \text{d}M \frac{\text{d}n(M, z)}{\text{ d}M}\langle N_g \rangle_{M} b_{\text{lin}}(M, z).
\end{equation}
The Fourier transform $u_{g}(k; M, z)$ is formally calculated for subhalos of galaxies included in the parent halo. However, the expression $u_{g}(k; M, z)$ can be replaced by $u_{\text{dm}}(k; M, z)$ defined by the formulas~(\ref{eq:normfourie}),~(\ref{eq:fourie}) with a reasonable accuracy~\cite{10.1046/j.1365-8711.2001.04202.x,Sheth:2002cs}. The exponent $p$ depends on the average number of galaxies in the halo $\langle N_{g} \rangle$ and on the type of its constituent galaxies~\cite{Cooray:2002dia}.

The average number of galaxies $\langle N_{g}\rangle_{M}$ in a halo of mass $M$ is represented as the sum of galaxies located in the center of a halo of mass $M$ and satellite galaxies distributed inside this halo~\cite{Zheng:2004id}
\begin{equation}
    \langle N_g \rangle_{M} = \langle N_{\text{cen}} \rangle_{M} + \langle N_{\text{sat}} \rangle_{M}.
\end{equation}
Assuming that the random variable $N_g$ has a Poisson distribution, the average number of pairs of galaxies in the halo $\langle N_{g}(N_{g} - 1)\rangle_{M}$ can be represented as
\begin{equation}
\langle N_{g}(N_{g} - 1)\rangle_{M} = \langle N_{\text{cen}}(N_{\text{cen}} - 1)\rangle_{M} + 2\langle N_{\text{cen}}N_{\text{sat}}\rangle_{M} + \langle N_{\text{sat}}(N_{\text{sat}} - 1)\rangle_{M} ,
\end{equation}
where, due to the conditions, $N_{\text{sat}} = 0$ for $N_{\text{cen}} = 0$ and $N_{\text{sat}} > 0$ for $N_{\text{cen }} = 1$, and $\langle (N_{g} - \langle N_{g} \rangle)^2 \rangle_{M} = \langle N_{g} \rangle_{M}$, we have $\langle N_{\text{cen}}N_{\text{sat}}\rangle_{M} = \langle N_{\text{sat}}\rangle_{M}$, $\langle N_{\text{sat} }(N_{\text{sat}} - 1)\rangle_{M} = \langle N_{\text{sat}}\rangle_{M}^2$.
Thus, the expression for the moment of the second order takes the form~\cite{Zheng:2004id}
\begin{equation}
    \langle N_{g}(N_{g} - 1)\rangle_{M} = 2\langle N_{\text{sat}}\rangle_{M} + \langle N_{\text{sat}}\rangle_{ M}^{2}.
\end{equation}
 With this in mind, the expressions included in (\ref{eq:3dps_gg_1h}, \ref{eq:3dps_gg_2h}) are written as
\begin{equation}\label{eq:N_mult_u2}
    \langle N_{g}(N_{g} - 1)\rangle_{M}|u_{g}(k; M, z)|^p = 2\langle N_{\text{sat}}\rangle_{M }|u_{g}(k; M, z)| + \langle N_{\text{sat}}\rangle_{M}^{2}|u_{g}(k; M, z)|^2,
\end{equation}
\begin{equation}\label{eq:N_mult_u}
    \langle N_g\rangle_{M}u_{g}(k; M, z) = \langle N_{\text{cen}}\rangle_{M} + \langle N_{\text{sat}}\rangle_{M }u_{g}(k; M, z),
\end{equation}
where, depending on $\langle N_{g} \rangle$, the galaxy located at the center of the halo make a single contribution to the dark matter density distribution spectrum, $u_{g}(k; M, z) = 1$~\cite{Cooray:2002dia}.

The average concentration of galaxies $\langle n_{g}(z) \rangle$ per unit comoving volume is given by
\begin{equation}
    \langle n_g(z) \rangle = \int \text{d} M \frac{\text{d} n(M, z)}{\text{d} M}\langle N_g\rangle_{M}.
\end{equation}
The quantities $\langle N_{\text{cen}} \rangle_{M}$, $\langle N_{\text{sat}} \rangle_{M}$ have the form~\cite{Zheng:2004id}
\begin{equation}
    \langle N_{\text{\text{cen}}}\rangle_{M} = \frac{1}{2}\left[1 + \text{erf}\left[\frac{\log M - \log M_{\text{min}}}{\sigma_{\log M}}\right]\right],
\end{equation}
\begin{equation}
     \langle N_{\text{\text{sat}}}\rangle_{M} = \left[\frac{M-M_0}{M_1}\right]^{\alpha},
\end{equation}
\begin{equation}
    \text{erf}(x) \equiv \frac{2}{\sqrt{\pi}}\int_{0}^{x}e^{-t^2}\text{d}t,
\end{equation}
where $M_{\text{min}}$ is the characteristic minimum mass of the halo containing the central galaxy, $M_{\text{0}}$ is the threshold mass at which there are no satellites in the halo, $M_{\text{1} }$ is the characteristic mass of the halo for which, on average, there is at least one satellite in the halo under the condition $M-M_{\text{0}} \geq M_{\text{1}}$, $\sigma_{\log M }$ is the halo mass variance due to the mass -- luminosity ratio ($M/L$), $\alpha$ is the slope of the power law in the distribution of the number of satellites. In our analysis, we set $M-M_{\text{0}} = M-M_{\text{min}}$. This condition means that satellite galaxies are presented only in the halo containing the central galaxy. The numerical values of the parameters used in our analysis are presented in Table~\ref{tab:hod_pars}
\begin{table}[htb]
\small
\caption{
The values of the parameters used in numerical analysis~\cite{Zheng:2004id}.}
\label{tab:hod_pars}
\begin{center}
\begin{tabular*}{\textwidth}{c@{\extracolsep{\fill}}cccc}
\hline
\hline
$\log(M_{\text{min}}/M_{\text{sun}})$ & $\log (M_{\text{0}}/M_{\text{sun}})$ & $\log (M_{\text{1}}/M_{\text{sun}})$ & $\sigma_{\log M}$ & $\alpha$ \\
\hline
11.68 & 11.86 & 13.00 & 0.15 & 1.02 \\
\hline
\end{tabular*}
\end{center}
\end{table}

The window function $W_{g}(z)$ characterizes the distribution of the number of galaxies in a given catalog over the redshift. In this work, we use the 2MASS Redshift Survey (2MRS) catalog, which includes more than 43,000 objects, with redshifts up to $z \sim 0.1$ and a sky coverage fraction of $f_{sky} \approx 0.91$~\cite{ Huchra_2012}. As indicated in~\cite{Ando:2017wff}, the differential distribution of the number of galaxies in this catalog is described with a good degree of accuracy by the following parameterization
\begin{equation}
    \frac{\text{d}N_{\text{2MRS}}}{\text{d}z} = \frac{N_{\text{2MRS}}\beta}{\Gamma[(m+1)/ \beta]}\left(\frac{1}{z_0}\right)\left(\frac{z}{z_0}\right)^m\exp{\left[-\left(\frac{z}{ z_0}\right)^{\beta}\right]},
\end{equation}
where $\Gamma[(m+1)/\beta]$ is the gamma function, $N_{\text{2MRS}} = 43182$, $\beta = 1.64$, $m = 1.31$, $z_0 = 0.0266$. Thus, the window function $W_{g}(z)$, normalized to unity, takes the form
\begin{equation}
    W_{g}(z) = \frac{\text{d}z}{\text{d}\chi} \left[\frac{1}{N_{\text{2MRS}}} \frac{\text {d}N_{\text{2MRS}}}{\text{d}z}\right].
\end{equation}
\section{Cross-correlation angular power spectrum of dark matter and galaxies}\label{sec:cross_corr_dmgal}
The cross-correlation angular power spectrum of dark matter and the distribution of the number of galaxies for a given catalog in Limber approximation~\cite{1953ApJ...117..134L} is written as~\cite{Cooray:2002dia, Fornengo:2013rga}
\begin{equation}
   \label{eq:aps_dmgal}
    C_l^{dm, g} = \int_{0}^{\infty} \
    \frac{\text{d}\chi}{\chi^2}
    \overline{W}_{dm}(\chi)W_{g}(\chi)\
    P_{dm, g}\left(k=\frac{l}{\chi}, z\right),
\end{equation}
where the nonlinear cross-correlation power spectrum $P_{dm, g}\left(k, z\right)$ has the following form
\begin{equation}
     P_{\text{dm, g}}(k, z) = P_{\text{dm, g}}^{1h}(k, z) + P_{\text{dm, g}}^{2h} (k, z),
\end{equation}
\begin{equation}
    P_{\text{dm, g}}^{1h}(k, z) = \int \text{d}M \frac{\text{d}n(M, z)}{\text{d}M }\frac{\langle N_g\rangle_{M}}{\langle n_g(z) \rangle}u_{\text{g}}(k; M, z)\left(\frac{M}{\overline{ \rho}(0)}\right)u_{\text{dm}}(k; M, z),
\end{equation}
\begin{align}
    \nonumber P_{\text{dm, g}}^{2h}(k, z) & = \left[\int \text{d}M \frac{\text{d}n(M, z)}{\text {d}M}b_{\text{lin}}(M; z)\left(\frac{M}{\overline{\rho}(0)}\right)u_{\text{dm}}(k; M, z)\right]\\
    & \times \left[\int \text{d}M \frac{\text{d}n(M, z)}{\text{d}M}b_{\text{lin}}(M; z) \frac{\langle N_g\rangle_{M}}{\langle n_g(z) \rangle}u_{\text{g}}(k; M, z)\right] P_{\text{lin}}(k , z).
\end{align}
The expression for the power spectrum $P_{dm, g}\left(k, z\right)$ on large scales can be approximately expressed in terms of the galactic bias as $P_{dm, g}\left(k, z\right) \approx b_ {g}(z) P_{dm, dm}\left(k, z\right)$~\cite{Seljak:2000gq}.

\section{Experimental Setup}\label{sec:exp_setup}

the Spectr-Roentegn-Gamma (SRG)~\cite{Sunyaev:2021bln} is the Russian X-ray observatory created with the participation of Germany and launched in July 2019. This orbital observatory is designed for deep observation and mapping of the Universe in a wide range of energies. The SRG has two telescopes eROSITA~\cite{2010SPIE.7732E..0UP,2012arXiv1209.3114M,erosita2020} and ART-XC~\cite{2011SPIE.8147E..06P,Pavlinskypart1, Pavlinskypart2,2019ExA....48..233P,Pavlinsky:2021jsk, Pavlinsky:2021ynp}. Each of the telescopes has its own characteristics and is intended for observations in different energy ranges. The telescope mission consists of two parts: a survey mission lasting 4 years and point observations of selected sources lasting 2 years. The results obtained during this mission make it possible to obtain strong restrictions on the parameters of dark matter, which, in particular, can be detected in the decays of sterile neutrinos into X-ray photons.

 The eROSITA telescope has a field of view of $0.833$ deg$^2$, while the ART-XC has a field of view of about $0.3$ deg$^2$. The last, however raises to $2$ deg$^2$ if consider ART-XC in concentrator mode, when total aperture flux of singly and doubly X-ray mirrors -reflected photons is detected without imaging~\cite{Pavlinskypart1, Pavlinskypart2, Pavlinsky:2021ynp}. Thus, the average exposure time for one year of observations in this field of view is 2500 seconds and 6100 seconds, respectively. One of the most important parameters that determine the quality of observations is the value of grasp. Grasp is defined as the product of the effective area of the telescope and the field of view. This parameter is responsible for the quality of observations when covering a certain part of the sky~\cite{Predehl:2020waz,Pavlinsky:2021ynp}. The angular and energy resolution of the telescopes are presented in summary Table~\ref{tab:ttp}.
\begin{table}[!htb]
  \caption{Telescopes Technical Performance.}
  \label{tab:ttp}
  \begin{center}
    \begin{tabular*}{\textwidth}{l@{\extracolsep{\fill}}cc}
    \hline
    \hline
      {} & eROSITA\ & ART-XC\ \\
      \hline
      energy range [keV] & 0.2 -- 10 & 4 -- 30\\
      energy resolution & 138\,eV at 6\,keV & 10\% at 14\,keV \\
      field of view (FOV) [deg$^2$] & 0.833 & 0.3 -- 2.0 \footnote{FOV [deg$^2$]: Telescope 0.3, Concentrator 1.7, Full 2.0 \cite{Pavlinskypart1, Pavlinskypart2, 2019ExA....48..233P}.}\\
      \hline
    \end{tabular*}
  \end{center}
\end{table}

For reliable detection of a signal from dark matter decays, a good knowledge of the background is required. Estimates of the expected background for the SRG telescopes are presented in~\cite{Predehl:2020waz, Pavlinsky:2021jsk, Pavlinsky:2021ynp, Barinov:2020hiq}, and in this paper we follow the methodology described in~\cite{Barinov:2020hiq}. Collected background data show that the background, like the rest of the telescope parameters, remains very stable~\cite{Pavlinskypart1, Pavlinskypart2, Pavlinsky:2021jsk, Pavlinsky:2021ynp}.

\section{Analysis Procedure}\label{sec:space_cons}
In the previous Sections, the general formalism for calculating the angular power spectra has been outlined. In this Section, we describe a statistical analysis procedure for obtaining constraints in the space of sterile neutrino parameters and analyze the resulting constraints.

The calculation of the angular power spectra is performed exactly according to the formalism from Sections~\ref{sec:formalism},\ref{sec:cross_corr_dmgal}. The \texttt{HALOMODEL}~\cite{halomodel}, \texttt{HMF}~\cite{hmf}, \texttt{COLOSSUS}~\cite{colossus} packages, as well as the \texttt{CAMB}~\cite{camb} package are used to calculate the linear matter power spectrum $P_{\text{lin} }(k, z)$, the halo mass function $\text{d}n/\text{d}M(M,z)$, the linear bias $b_{i}(M; z)$ and the Fourier transform of the dark matter density distribution function $u_{i}(k; z, M)$. The following $\Lambda$CDM parameters were used in the calculation~\cite{ParticleDataGroup:2022pth}: $h=0.674, \Omega_{\Lambda}=0.685, \Omega_m=0.315, \Omega_b=0.0493, \Omega_{\text{CDM}}=0.265, \sigma_8=0.811$. To speed up the process of calculating the angular power spectra, we first calculate the linear matter power spectrum $P_{\text{lin} }(k, z)$, the halo mass function $\text{d}n/\text{d}M(M,z)$, the linear bias $b_{i}(M; z)$ and the Fourier transform of the dark matter density distribution function $u_{i}(k; z, M)$ for a given uniform grid $k, z$ and $M$. Next, these datasets are cached and then used as anchor points for fast interpolation of data in the required range. The quality of the interpolation is additionally tested on a random sample of variables $k, z, M$, in order to make sure that there are no artifacts and outliers. The interpolation error is observed of the order of several percent, which is completely insignificant within the framework of the general uncertainties of the analysis~\cite{Cooray:2002dia}. Interpolation functions for the $P_{\text{lin} }(k, z)$, $\text{d}n/\text{d}M(M,z)$, $b_{i}(M; z)$ and $u_{i}(k; z, M)$ are obtained, the power spectra are calculated using formulas~(\ref{eq:1hterm}),~(\ref{eq:2hterm}). The integrals included in expressions~(\ref{eq:1hterm}),~(\ref{eq:2hterm}) are calculated numerically by two different methods. First, the integrals are calculated using the Simpson method, then the integration is carried out using the Vegas Monte Carlo method. The number of integration samples is increased iteratively in such a way that the calculation results turn out to be stable and the calculation error become constant. The integration limits in expressions~(\ref{eq:1hterm}),~(\ref{eq:2hterm}) and~(\ref{eq:aps}) are chosen to be $z: [0.01, 0.2], M: [10^6, 10^{16}] M_{sun}$. The procedure for calculating the power spectra in the extended Limber approximation is carried out in a completely similar way as in Limber approximation up to the replacement of $l$ by $l + 1/2$. 

\begin{sidewaysfigure}[p]
\centering
\includegraphics[width=1.0\linewidth]{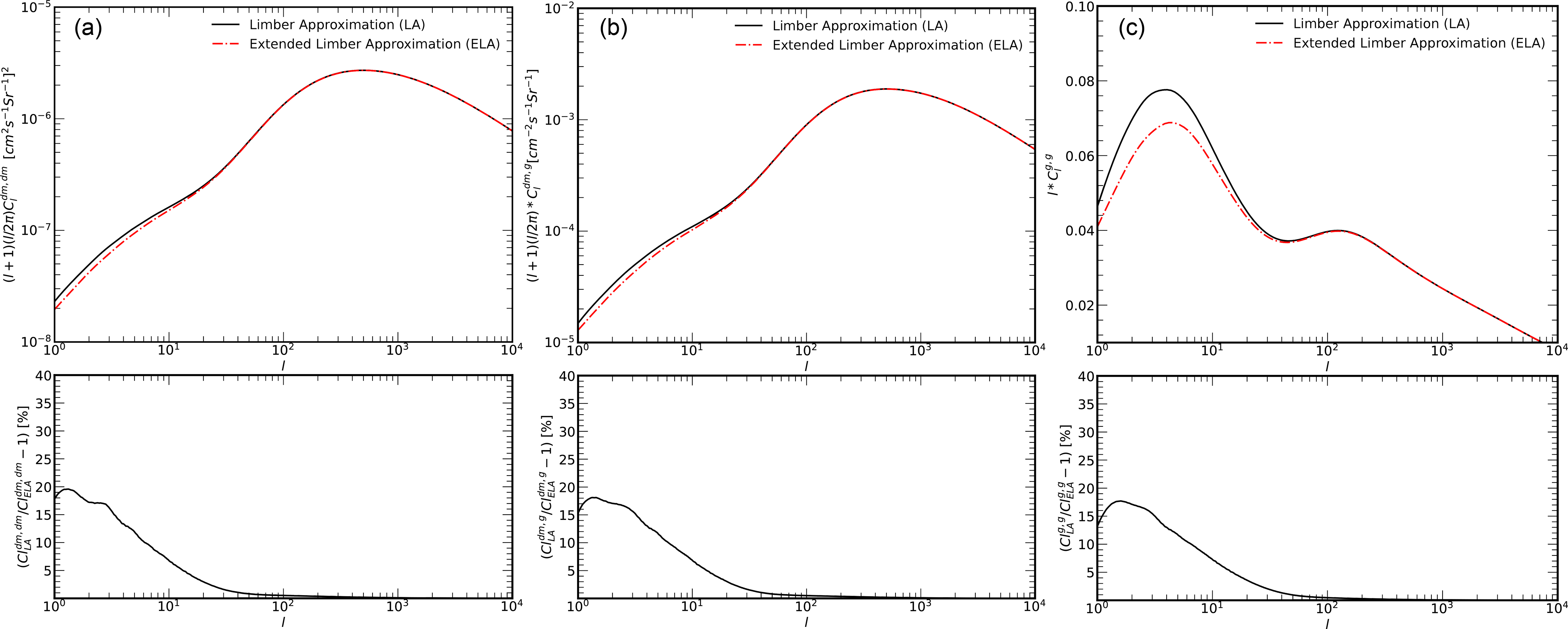}
\caption{Angular power spectra are constructed for eROSITA energy resolution $\sigma_{E}$. Panel (a): Autocorrelation angular power spectrum of dark matter due to decays of dark matter particles. Panel (b): Cross-correlation angular power spectrum of dark matter and galaxies, constructed for the 2MRS catalog using the HOD formalism. This spectrum constructed within the $P_{dm, g}\left(k, z\right) \approx b_{g}(z) P_{dm, dm}\left(k, z\right)$ approximation. Panel (c): Autocorrelation angular power spectrum of galaxies constructed for the 2MRS catalog using the HOD formalism. This spectra are constructed for the sterile neutrino parameters ($m_{\nu_{s}}=7.12 \hspace{0.2pt} \text{keV}, \sin^2(2\theta) = 7.6\times10^{-11 }$), where the energy range for the average X-ray photon emission intensity is $\left[3.4, 3.6\right] \text{keV}$ for reference point adopted in~\cite{Zandanel:2015xca}. The bottom panels show residuals between Limber approximation and extended Limber approximation.}
\label{fig:corr_specta_erosita}
\end{sidewaysfigure}

Figure~\ref{fig:corr_specta_erosita} shows the results of calculations of the correlation angular power spectra for all pairs of signatures for the eRosita telescope. The spectra were calculated both within the Limber approximation and within the extended Limber approximation. The bottom panel of Figure~\ref{fig:corr_specta_erosita} shows the relative difference between the correlation spectra calculated for both approaches. As can be seen, the relative difference between the spectra rapidly decreases with increasing multipole value from $20 \%$ for $l \sim 1 - 2$ to $7 \%$ for $l \sim 10$. For multipole values $l \sim 50$ and above, the difference becomes negligible. It is worth noting that the shape of the spectra calculated in both approaches for small multipoles is preserved, changing the advantage only in amplitude. Thus, we can be sure that for multipoles $l \geq 50$, the Limber approximation gives the same results for all pairs of signatures as the extended Limber approximation. For multipoles $l \sim 1 - 10$, the use of the extended Limber approximation becomes justified.

In order to determine the constraints on the parameters of sterile neutrinos, we use the standard $\chi^2$ analysis procedure
\begin{equation}
    \chi^2 = \sum_{l,l',E,E'}\left(C_{l,E}^{\text{dm,g}} - C_{l,E}^{\text{m}}\right)\left(\delta C_{l,l',E,E'}^{2}\right)^{-1}\left(C_{l',E'}^{\text{dm,g}} - C_{l',E'}^{\text{m}}\right),
\end{equation}
where we require the value $\chi^2 = 2.71$, which corresponds to the $2\sigma$ or $95\%$ C.L. for one-sided test with one degree of freedom ($\sin^2(2\theta) $)~\cite{ParticleDataGroup:2022pth}. $C_{l,E}^{\text{dm,g}}$ is the theoretical value of the cross-correlation angular power spectrum for a given multipole $l$ with a given photon energy $E$. $C_{l,E}^{\text{m}}$ is the measured cross-correlation spectrum, $\delta C_{l,l',E,E'}$ is the uncertainty covariance matrix, which can be computed as~ \cite{Zhang:2004tj, Cuoco:2006tr,2009MNRAS.400.2122A, Ando:2014aoa,Campbell:2014mpa}
\begin{equation}
\label{eq:delta_Cl}
    \delta C_{l,l',E,E'}^{2} = \frac{\delta_{l,l'}}{(l + 1/2)l f_{sky}}\left(C_{l,E}^{\text{dm,g}}C_{l',E'}^{\text{dm,g}} + \left(C_{l,E, E'}^{\text{dm,dm}} + C_{N,E}^{\gamma}\delta_{E,E'}\right)\left( C_{l'}^{\text{g,g}}+C_{N}^{\text{g}}\right)\right).
\end{equation}

The quantity $C_{N,E}^{\gamma}$  is the shot noise due to X-ray photons
\begin{equation}\label{eq:CNE}
    C_{N,E}^{\gamma} = \frac{4 \pi f_{sky}}{W_{l}^2}\frac{\overline{I_{\gamma}}^2}{N_{E}},
\end{equation}
where $f_{sky}$ is the part of the sky covered by observations, $W_{l} \equiv \exp{\left(-l^2\sigma_b^2 /2\right)}$ is the point spread function~\cite{Zhang:2004tj, Cuoco:2006tr}, $\sigma_b$ is the angular resolution of the telescope (in arcsec), $\overline{I_{\gamma}}$ is the sky-averaged total X-ray intensity in the given energy range. It is determined by the sum of several contributions: the extragalactic radiation of X-ray photons, the radiation of the galaxy, and the instrumental background. The sky-averaged radiation intensity of X-ray photons due to the decay of dark matter particles in a given energy range is calculated as
\begin{equation}
     \overline{I_{dm}} = \int_{E_{min}}^{E_{max}}\text{d}E I_{\text{dm}}(E),
\end{equation}
where the integrand is the intensity of X-ray photons
\begin{equation}
    I_{\text{dm}}(E) = \int_{0}^{\infty}\text{d}\chi W_{dm}(E, z).
\end{equation}
The contribution of the galaxy can be modeled in various ways, but at energies above $1$ keV it is insignificant~\cite{Barinov:2020hiq}. The instrumental background in this case is the particle background for the X-ray telescope. For the interesting energies it can be treated as constant with a good degree of accuracy~\cite{Predehl:2020waz, Pavlinsky:2021jsk, Pavlinsky:2021ynp}. The quantity $N_{E}$ included in~(\ref{eq:CNE}) is the number of registered X-ray photons
\begin{equation}
    N_{E} = T_{obs} \Omega_{\text{FOV}} f_{sky} \int_{E_{min}}^{E_{max}}\text{d}E I_{\gamma}(E) A_{eff}(E),
\end{equation}
where $T_{obs}$ is the observation time, $\Omega_{\text{FOV}}$ is the field of view of the telescope, $A_{eff}(E)$ is the effective area of the telescope including vignetting effects. $C_{N}^{g}$ is the shot noise of galaxies
\begin{equation}
    C_{N}^{g} = \frac{4 \pi f_{sky}}{N_{g}},
\end{equation}
where $N_{g}$ is the number of galaxies in the given catalog.
\section{Results}\label{sec:results}
\begin{figure}[!h]
\centering
\includegraphics[width=1.0\linewidth, height=0.5\textheight]{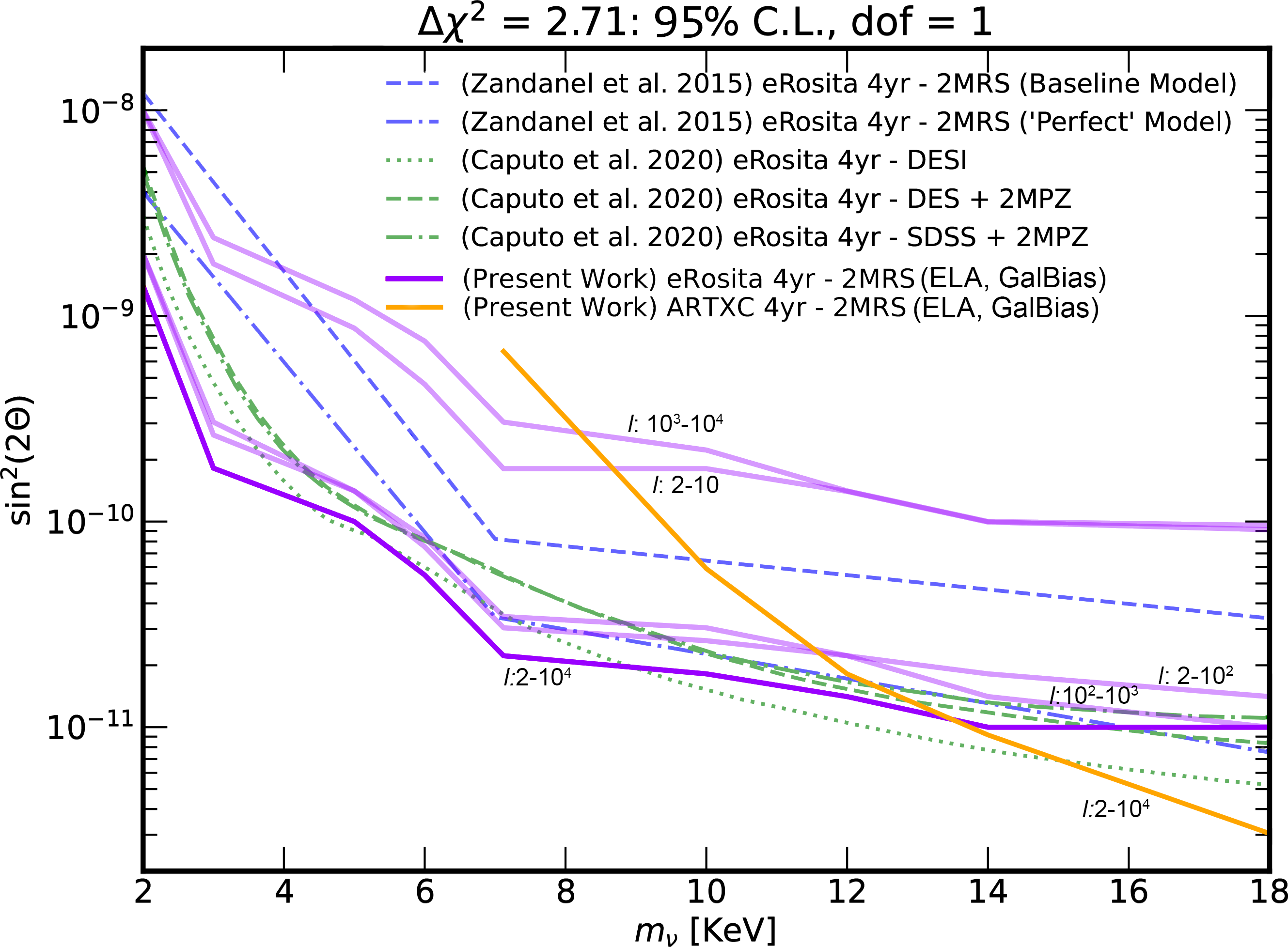}
\caption{Expected constraints on the parameters of sterile neutrinos obtained in the framework of our analysis for various ranges of multipoles. The purple line corresponds to the constraints for the eROSITA telescope (the translucent purple lines show the contributions to the constraints for different ranges of multipoles). The yellow line corresponds to the ART-XC telescope. For comparison, the constraints from the works~\cite{Zandanel:2015xca,Caputo:2019djj} are presented. The observation time is 4 years in the full sky survey mode.}
\label{fig:cosmo_cons}
\end{figure}

\begin{figure}[!h]
\centering
\includegraphics[width=1.0\linewidth, height=0.5\textheight]{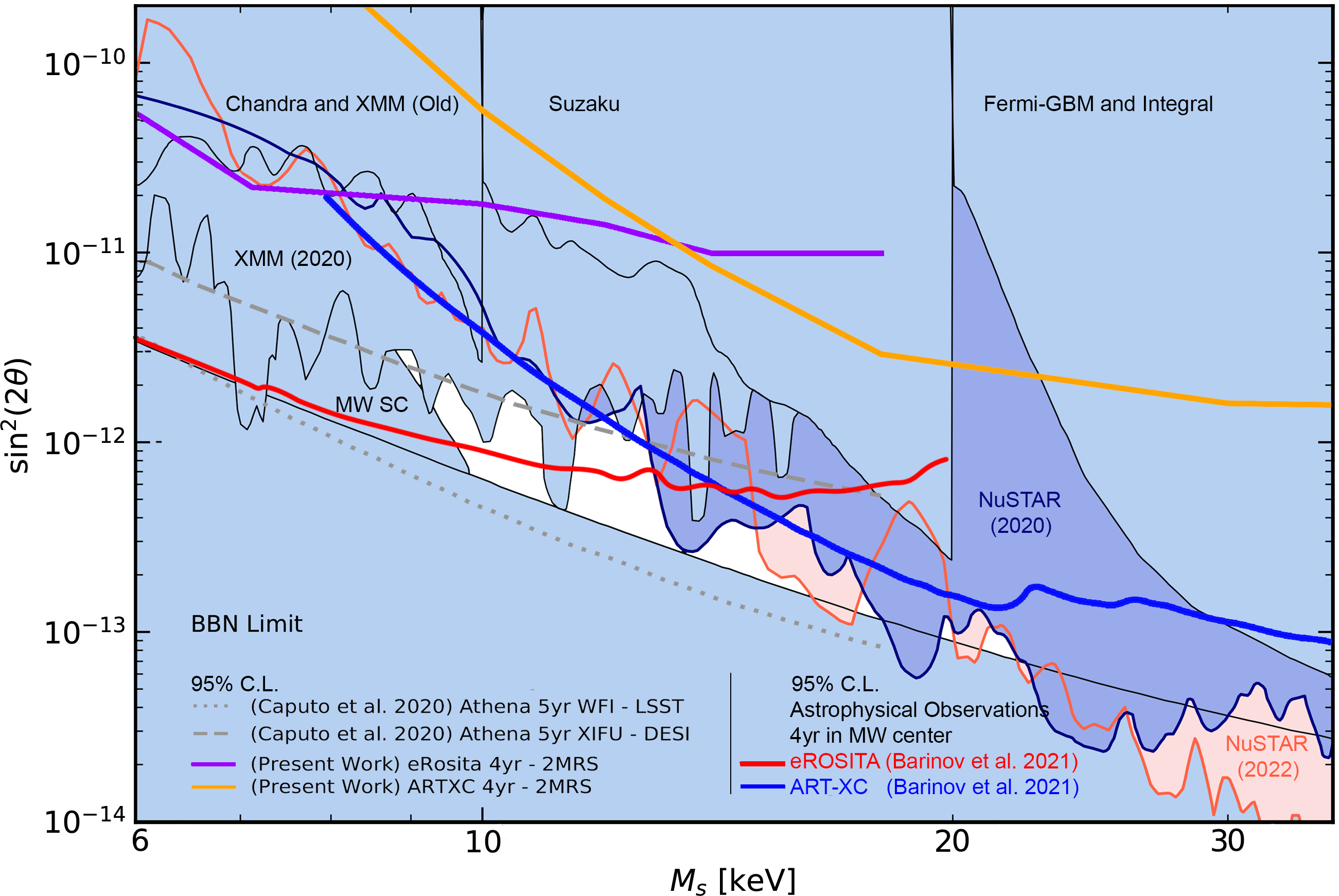}
\caption{Expected Constraints on the parameters of sterile neutrinos obtained in our analysis (same as Figure~\ref{fig:cosmo_cons}) in comparison with the constraints that can be obtained in the framework of the astrophysical observations of the center of the Milky Way in a cone with an opening angle of 60 degrees~\cite{Barinov:2020hiq} (the red line corresponds to the constraints for the eROSITA telescope, the blue line corresponds to the ART-XC telescope). Background Corrected means that the background has been additionally normalized taking into account additional features for the ART-XC telescope~\cite{Pavlinsky:2021ynp} compared to the previous work~\cite{Barinov:2020hiq}. Additionally, we present the constraints that can be obtained for the Athena telescope~\cite{Caputo:2019djj}.}
\label{fig:astro_cosmo_cons}
\end{figure}

We obtain the expected constraints presented in Fig.~\ref{fig:cosmo_cons} and Fig.~\ref{fig:astro_cosmo_cons}. Figure~\ref{fig:cosmo_cons} shows the constraints to be obtained for the eROSITA telescope and the ART-XC telescope in the 4-year all-sky survey mode. As can be seen from the presented results, our constraints are in good agreement with the constraints presented in~\cite{Zandanel:2015xca,Caputo:2019djj}. Note that in our work we calculate the correlation angular power spectra not limited to the Limber approximation, but also using the extended Limber approximation. Despite the fact that for multipoles $l > 100$ the power spectra differ negligibly, we see that the contribution of small multipoles is noticeable at $l = 2 - 50$ and also contributes to the resulting constraints. Therefore, taking into account small multipoles turns out to be relevant when studying spectra on large scales (small multipoles). In Fig.~\ref{fig:cosmo_cons}, we additionally illustrate the contribution of the different multipoles to the resulting constraints. As follows from Fig.~\ref{fig:cosmo_cons}, the most significant constraints are associated with multipoles $l \sim 10^2$--$10^3$ and $l \sim 10^1$--$10^2$. It is worth additionally noting that the constraints are also sensitive to splitting into multipole bins.

\newpage
Figure~\ref{fig:astro_cosmo_cons} shows the modern constraints on the parameters of sterile neutrinos and the constraints obtained in the framework of our analysis (see Fig.~\ref{fig:cosmo_cons}), as well as the constraints that can be obtained in the framework of astrophysical observations for the SRG mission when observing the center of the Milky Way in the cone with an opening angle of 60 degrees. In this Figure, we additionally present the constraints that can be obtained for the Athena telescope presented in~\cite{Caputo:2019djj}. 

As follows from the analysis data, the limits within the cross-correlation analysis are weaker than the limits that can be obtained within the framework of the local astrophysical observations. A comparison of the limits for both approaches shows that they generally agree well with the limits obtained from previous observations such as Chandra, XMM, Suzaku, Fermi-GBM and Integral. The NuSTAR data presented in Figure~\ref{fig:astro_cosmo_cons}, however, turns out to be even stronger. However, it should be noted that the limitations obtained from astrophysical observations largely depend on the chosen observation strategy and are rather sensitive to models of dark matter density distribution.
\section{Summary}
\label{sec:conclusion}
In this work we investigated auto and cross-correlation power spectra for decaying dark matter in the model with decays of sterile neutrinos into active neutrinos and photons, as well as power spectra for the 2MRS catalog of galaxies. We considered the procedure of auto and cross-correlation analysis within the framework of the approach described in works~\cite{Zandanel:2015xca, Caputo:2019djj}, and also carried out the analysis using the extended Limber approximation~\cite{LoVerde:2008re}. This approach makes it possible to calculate the angular power spectra at small multipoles, which is of interest for studying dark matter traces on large cosmological scales. It was shown that on medium and large multipoles the Limber approximation is very efficient.

We have analyzed the constraints on the parameters of sterile neutrinos for both the eROSITA telescope and the ART-XC telescope and comparison with modern constraints is carried out. As part of the work, it was shown that the results obtained by us for the eROSITA telescope are in good agreement with the results of previous works~\cite{Zandanel:2015xca, Caputo:2019djj}. However, in our analysis, we used the refined data on the X-ray background~\cite{Predehl:2020waz, Pavlinsky:2021jsk, Pavlinsky:2021ynp, Barinov:2020hiq}.

The analysis was carried out for the first time for the ART-XC telescope. It is shown that the ART-XC telescope has a good potential for probing models with decaying sterile neutrinos in the high-energy region. The strongest constraints are collected from the region of medium multipoles.

It should be noted that the constraints on the parameters of sterile neutrinos obtained in the framework of this paper and in Refs~\cite{Zandanel:2015xca, Caputo:2019djj} for SRG telescope are weaker than the constraints that can be obtained in the framework of the analysis of the local astrophysical observations~\cite{Perez:2016tcq,Barinov:2020hiq}. However, these limits appear to be comparable to those obtained from past X-ray observations such as Chandra, XMM, Suzaku, Fermi-GBM and Integral. It is worth noting that in~\cite{Caputo:2019djj} the constraints that can be obtained within the framework of the Athena project are additionally presented. These limits have a very high potential, as they are comparable with modern limits obtained from astrophysical observations.

We conclude that the allowed region of the dark matter parameters in models with decaying sterile neutrinos can be additionally limited not only from the results of previous astrophysical observations but also independently within the framework of the correlation analysis.

\acknowledgments
The author is grateful to D. Gorbunov for the suggestion to consider this problem. The author is also grateful to R. Burenin, R. Krivonos, and A. Caputo for discussions and important comments. This work was supported by the Russian Science Foundation (RSF) grant 22-12-00271.

\bibliographystyle{apsrev}
\bibliography{bibliography}

\end{document}